\documentclass[12pt]{article}

\usepackage{amssymb}
\usepackage{amsmath,bm}
\usepackage{amsthm}
\usepackage{geometry}
\usepackage{graphicx}
\usepackage{apacite}
\usepackage{natbib}
\usepackage{hyperref}
\usepackage{parskip}
\usepackage{setspace}

\newtheorem{thm}{Theorem}

\newcommand{\blind}{0}

\begin{document}

\def\spacingset#1{\renewcommand{\baselinestretch}%
{#1}\small\normalsize} \spacingset{1}

\begin{titlepage}

\if0\blind
{
\title{\bf Fast Bayesian Record Linkage With Record-Specific Disagreement Parameters \footnotetext{The author thanks Jiaying Gu and Shari Eli, as well as Martin Burda, En Hua Hu and others for their help and suggestions. The comments of two anonymous referees were extremely helpful. This research was supported by the Social Sciences and Humanities Research Council of Canada, and by NIH grant number P01 AG10120, Early Indicators of Later Work Levels, Disease and Death, Dora L. Costa, principal investigator.}}
\author{Thomas Stringham\thanks{Department of Economics, University of Toronto, 150 St. George Street, Toronto, Ontario, M5S 3G7, Canada. Email: \href{mailto:tom.stringham@mail.toronto.ca}{tom.stringham@mail.utoronto.ca}} \smallskip
\\ Department of Economics, University of Toronto}
\date{\today}
\maketitle
} \fi

\if1\blind
{
\title{\bf Fast Bayesian Record Linkage With Record-Specific Disagreement Parameters}
\date{\today}
\maketitle
}
\fi

\vskip 2.5em
\begin{abstract}
\noindent Researchers are often interested in linking individuals between two datasets that lack a common unique identifier. Matching procedures often struggle to match records with common names, birthplaces or other field values. Computational feasibility is also a challenge, particularly when linking large datasets. We develop a Bayesian method for automated probabilistic record linkage and show it recovers more than 50\% more true matches, holding accuracy constant, than comparable methods in a matching of military recruitment data to the 1900 US Census for which expert-labelled matches are available. Our approach, which builds on a recent state-of-the-art Bayesian method, refines the modelling of comparison data, allowing disagreement probability parameters conditional on non-match status to be record-specific in the smaller of the two datasets. This flexibility significantly improves matching when many records share common field values. We show that our method is computationally feasible in practice, despite the added complexity, with an R/C++ implementation that achieves significant improvement in speed over comparable recent methods. We also suggest a lightweight method for treatment of very common names and show how to estimate true positive rate and positive predictive value when true match status is unavailable. \\
\vspace{0in}

\bigskip
\end{abstract}

\setcounter{page}{0}
\thispagestyle{empty}
\end{titlepage}
\pagebreak \newpage
\spacingset{1.45}

\section{Introduction}

The record linkage problem of identifying unique individuals between two datasets without common unique identifiers arises in various applied settings. Typically, linkage, also called matching, is performed to create a new, combined dataset that can be used to answer economic and policy questions. In an example from the economics literature, \cite{costa2013} uses recruitment data from veterans of the US Civil War linked to multiple US censuses to examine life outcomes associated with leadership roles during war.

Recent approaches to the record linkage problem in the statistical literature include the Bayesian method (Beta Record Linkage, or $\beta$RL) of \cite{sad2017}, of which our proposed model is a refinement. $\beta$RL is itself a robust, high-performance refinement of previous Bayesian methods including \cite{for2001} and \cite{lar2002, lars2005, lars2010}.

Bayesian methods perform well but are computationally complex, typically running more slowly than comparable frequentist methods. \cite{mcv2017} propose improving speed through a two-step approach involving a post hoc blocking procedure (see \cite{christen2011} for a general discussion of blocking). \cite{imai2019} use parallelization across processor cores to develop a very fast frequentist package, {\tt fastLink}, and also offer a Bayesian variant, though the bipartite restriction is not enforced in the model. \cite{marchant2021} fully endogenize the blocking procedure in large datasets and use modern parallel computing techniques to improve scalability.

Record linkage problems frequently require dealing with what we refer to as the common name problem, which is that many records (usually people) share very common names, birthplaces or other field values and are harder to link than records with uncommon names and field values. \cite{newc1959}, in an early example of computer record linkage, addressed this problem by manually constructing probability weights to adjust for frequency of common names. \cite{wink1988, wink1989} and \cite{yancey2000} proposed frequency-based weights for modern frequentist matching procedures. {\tt fastLink} includes an option for ex-post reweighting by first name frequency. Alternative string distance metrics that take frequency into account, such as the ``soft" TF-IDF measure of \cite{cohen2003}, have also been proposed for dealing with the issue as it applies to character fields like first and last name, though this imposes considerable computational cost in large matchings.

In this paper, we outline a statistical model for record linkage which allows for record-specific heterogeneity, a refinement of the model of \cite{sad2017}. In particular, disagreement probability parameters conditional on non-match status are allowed to vary by record in the smaller of the two datasets, mitigating the common name problem by giving the model flexibility over every field value that appears in the data. In an application, we link recruitment data from veterans of the Union Army in the US Civil War to full count 1900 US Census data. ``True" matches, made manually by experts using additional information, are used to evaluate performance. Our significant relaxation of the model, which adds computational expense, leads to identifying more than 50\% more true matches than in a baseline Bayesian model without record-specific disagreement parameters, when holding accuracy constant.

In addition, we propose an alternative, lightweight method for treatment of very common names. We modify comparison data to include an indicator for very common names that otherwise spuriously produce exact matches in record pairs. This modification slightly improves matching performance in the Union Army application, relative to a baseline model, and adds very little computational expense. It is useful when other methods for dealing with common names, such as record-specific parameters, are computationally prohibitive. The method can be used in combination with record-specific parameters, though we find it does not further improve matching performance in that case.

We implement a version of the Gibbs sampler of \cite{sad2017} to draw from the posterior, achieving much higher computation speed than in the original implementation, mainly by drastically reducing memory requirements related to comparison data (see Section \ref{imp}). We also achieve speed comparable to that of frequentist methods, even when record-specific parameters are used, suggesting that estimating large matchings using a fully Bayesian approach that enforces a bipartite matching and allows record-specific parameters can be practical using commonly available computational resources.

\section{The record linkage problem}
\subsection{The Fellegi-Sunter framework}

The framework for probabilistic record linkage in \cite{fs1969} is the foundation for all subsequent work in the record linkage literature, and we briefly describe it here.

Two datafiles, labeled $\textbf{A}$ and $\textbf{B}$, contain records indexed by $i$ and $j$ respectively and are measured from a common underlying population. $\textbf{A} \times \textbf{B}$ contains all pairs of records where the first record is in $\textbf{A}$ and the second in $\textbf{B}$, and can be partitioned into disjoint sets $\textbf{M}$ and $\textbf{U}$, where $\textbf{M}$ is the set of pairs that are true matches in the underlying population and $\textbf{U}$ the set of non-matches. A successful matching corresponds to identification of the elements of $\textbf{A} \times \textbf{B}$ that are in $\textbf{M}$ (this problem is trivial if a common unique identifier is available).

Records in $\textbf{A}$ and $\textbf{B}$ share descriptive characteristics or fields, indexed by $f=1, 2, \dots, F$, such as name, age and birthplace. For each field $f$ there are $L_f$ disagreement levels indexed by $l = 1, 2, ..., L_f$, where higher values indicate greater levels of disagreement. The comparison for a given record pair $(i,j) \in \textbf{A} \times \textbf{B}$ in field $f$, $\gamma_{ij}^f$, takes such a value $l$. A vector of comparisons for a given record pair $(i,j) \in \textbf{A} \times \textbf{B}$ is denoted $\gamma_{ij}$, while the matrix generated by stacking these vectors across all record pairs, called the comparison data, is denoted $\pmb{\gamma}(\textbf{A}, \textbf{B})$.

The number of records in $\textbf{A}$ is $n_A$, the number of records (rows) in $\textbf{B}$ is $n_B$ and $n_A \leq n_B$, without loss of generality. A toy record linkage problem is illustrated in Table \ref{tab:simple}, with $n_A=2$, $n_B=5$, and $F=3$.
\begin{table}[ht]
\centering
\begin{tabular}{ll}

\begin{tabular}{c|c|c|c} 
i&First & Last & Birth year \\ \hline
1&John & Lundrigan & 1848 \\ 
2&Jedediah & Smith & 1844 \\ 
\end{tabular}
\quad \quad

\begin{tabular}{c|c|c|c} 
j&First & Last & Birth year \\ \hline
1&John & Lundgren & 1848 \\
2&Jon & Lundregan & 1850 \\ 
3&Jedidiah & Smith & 1845 \\
4&John & Smith & 1844 \\ 
5&Jedediah & S & 1844 \\
\end{tabular}
\end{tabular}
\caption{Toy datafiles $\textbf{A}$ and $\textbf{B}$ to illustrate a census data application; not real data. The first column gives researcher labels, not identifiers.}
\label{tab:simple}
\end{table}

To produce an estimate of a matching using the comparison data, there must be a model for the comparison data specifying how likely some level of disagreement is for a particular field and a particular record pair. \cite{fs1969} assume that disagreement in comparison fields is independent conditional on match status (matched or unmatched). The probability of $\gamma_{ij}$ taking on disagreement level $l$ in field $f$, conditional on $(i,j) \in \textbf{M}$, is $m_{fl} = P(\gamma_{ij}^f = l | (i,j) \in \textbf{M}).$ Similarly, the probability of $\gamma_{ij}$ taking on disagreement level $l$ in field $f$, conditional on $(i,j) \in \textbf{U}$, is $u_{fl} = P(\gamma_{ij}^f = l | (i,j) \in \textbf{U}).$ The vector of probabilities $(m_{f1}, \dots m_{f{L_f}})$, which sums to 1, is denoted $m_f$ and the vector $(m_1, \dots m_F)$ is denoted $\mathbf{m}$. Similarly, the vector $(u_{f1}, \dots u_{f{L_f}})$ is denoted $u_f$ and the vector $(u_1, \dots u_F)$ is denoted $\mathbf{u}$.

Because $\textbf{M}$ and $\textbf{U}$ are not observed, the model is a latent class model, where $\textbf{M}$ and $\textbf{U}$ are the two classes. Given the conditional independence assumption we have, for a particular vector of comparisons $\gamma_{ij}^{obs}$ on the record pair $(i,j)$,
$$P(\gamma_{ij} = \gamma_{ij}^{obs} | (i,j) \in \textbf{M}) = \prod_{f=1}^F m_{f{l^{obs}}}^{\mathbf{1}((i,j) \in \textbf{M})} u_{f{l^{obs}}}^{\mathbf{1}((i,j) \in \textbf{U})} , $$
where $l^{obs}$, with some abuse of notation, indicates the observed disagreement level in each field. 

\sloppy
In \cite{fs1969}, possible matches are ranked by the log-likelihood ratio $P(\gamma_{ij} = \gamma_{ij}^{obs} | (i,j) \in \textbf{M}) / P(\gamma_{ij} = \gamma_{ij}^{obs} | (i,j) \in \textbf{U})$, and a thresholding decision rule, determined by a researcher-chosen target error rate, is used to arrive at a matching. The authors show this procedure yields optimal results for the specified error rates. This requires, however, a decision rule that allows for record pairs to be left unclassified, often an undesirable property.

\subsection{Further developments}

Most subsequent developments in probabilistic record linkage are extensions of the landmark work in \cite{fs1969}. \cite{wink1988} and \cite{jaro1989} proposed the use of an expectation-maximization (EM) algorithm to estimate conditional disagreement probability vectors $\mathbf{m}$ and $\mathbf{u}$. After an initial guess, the algorithm alternates between an expectation (E) step, where probabilities of each record pair being a match are computed, and a maximization step (M), in which $\mathbf{m}$ and $\mathbf{u}$ are re-estimated using MLE, until convergence. Mixture models estimated with EM have become the standard frequentist approach to record linkage \citep[see, e.g.,][]{sad2013, abr2019}. 

The Fellegi-Sunter probability model does not require each record in $\textbf{A}$ be matched to at most one record in $\textbf{B}$, and vice versa, though a final bipartite matching can be produced by a thresholding decision rule, as noted above. Indeed, the model assumes match status is independent across record pairs. As a result, it is possible and frequent for two mutually exclusive matches to have probabilities that sum to a number greater than one, suggesting the match status independence assumption is not met and the model is misspecified.

In a Bayesian setting, the matching itself is the parameter of interest, and a bipartite/one-to-one matching restriction can be enforced by restricting the support of the posterior. Bayesian methods also have the advantage of producing a full posterior which can be used in further inference. $\beta$RL, the method of \cite{sad2017}, built on \cite{for2001} and \cite{lar2002, lars2005, lars2010}, allowing for multiple levels of agreement and missing data.

\cite{mcv2017}, using the Bayesian model of \cite{sad2017}, outline a way to choose suitable small blocks using weights from a lightweight penalized likelihood procedure before estimating a full matching, in order to reduce computation time. In \cite{mcv20183}, the authors use this method to perform a matching of similar scale to the application in this paper. \cite{marchant2021} propose {\tt d-blink}, a comprehensive, scalable Bayesian procedure for record linkage and deduplication that builds on a literature (see \cite{tancredi2011, ste2015}) that models latent data directly.

In an alternative, frequentist approach, \cite{imai2019} develop a parallelized R package {\tt fastLink} for rapidly estimating matchings. They also include an option to include a Bayesian prior, but the bipartite restriction (one-to-one assignment) is not enforced.

\subsection{Blocking}

Because the number of comparisons that must be made rises quadratically with the number of records, since each record in $\textbf{A}$ must be compared with all records in $\textbf{B}$, researchers often ``block" datasets when matching large datasets, only comparing records that match exactly on one or more variables measured with low error. For example, if dealing with US census data, a researcher might only compare records in $\textbf{A}$ to those in $\textbf{B}$ with the same initials and state of birth, drastically reducing computational demands. With $b$ blocks of equal size, there are $b \times n_A/b \times n_B/b$, or $\frac{n_A n_B}{b}$ pairs to consider, instead of $n_A n_B$. However, blocking inevitably comes at the cost of discarding some true matches.

\subsection{Practical issues}
\label{prac}

For record pairs sharing common first or last names, exact agreement in one of the name fields is probable conditional on both match status and non-match status. In models with fixed disagreement parameters, this has the effect of lowering the probability of assigning a match in some record pairs that a human linker might readily mark as matched. 

The problem is aggravated within blocks with records dominated by common names. For example, within a ``W.S." block in US census data, a large proportion of records will be named William or surnamed Smith. There may be little separation between the two classes ($\textbf{M}$ and $\textbf{U}$), because of spurious exact matches on these common names, so that no matches can be identified within the block. Matchings in such blocks can be significantly improved by using the ``U-correction" procedure of \cite{mcv20183}, which adjusts the $u$ parameters to reflect the distribution of disagreements over all record pairs rather than only those in the block, and which we compare to our own method in Section \ref{ua}. This requires computation of marginal distributions of comparison data for all record pairs, however, which can add considerable computational expense.

A related problem occurs when one datafile is significantly larger than the other ($n_A << n_B$), and more generally when $n_B$ is very large, such that there are large numbers of observably identical or near-identical records in $\mathbf{B}$. The presence of false positives resulting from large $n_B$ tends to reduce both accuracy and the number of true matches recovered. Relatively small $n_A$ aggravates the problem as records missing in the smaller population cannot provide information that would help assign remaining records. Table \ref{smallna} illustrates this with a toy example.
\begin{table}[ht]
\centering
\begin{tabular}{c|c|c|c} 
\multicolumn{4}{c}{\textbf{A}} \\
i& First & Last & Birth yr \\ \hline
1& John & Lundrigan & 1848 \\ 
\emph{2} &\emph{Jonathan} &\emph{Lundregan} & \emph{1850} \\ 
\end{tabular}\quad \quad \begin{tabular}{c|c|c|c} 
\multicolumn{4}{c}{\textbf{B}} \\
j& First & Last & Birth yr \\ \hline
1&John & Lundgren & 1848 \\
2&Jon & Lundregan & 1850 \\ 
\end{tabular}
\caption{The small $n_A$ problem: the absence of the italicized record in $\textbf{A}$ makes less clear how to match record $i=1$.}
\label{smallna}
\end{table}
\cite{sad2017} discusses a related problem in which ``low overlap" of true matches between datafiles leads to poor performance in settings where $n_A \approx n_B$. In settings where $n_A << n_B$, the overlap of the two datafiles is necessarily small as a proportion of $n_B$.

This paper offers two new strategies for dealing with these problems. The first, which is computationally more expensive, involves estimating the disagreement probabilities conditional on non-match status, $\textbf{u}$, separately for each record in the smaller datafile, $\textbf{A}$---this is what is meant by ``record-specific" disagreement parameters.

This refinement increases the dimension of $\textbf{u}$ from $\sum_{f=1}^F L_f$ to $n_A \sum_{f=1}^F L_f$, since each record $i$ in $\textbf{A}$ is associated with its own parameters $u_{if} = (u_{if1}, \dots, u_{if{L_f}})$ for each field $f$. In the Union Army application discussed later on, this refinement significantly improves matching of records with less common names and birthplaces within all blocks and overcomes the problem of finding no matches in blocks with very common names. The increased dimensionality, however, makes matching expensive for blocks where the number of records in $\textbf{A}$ is very large. Record-specific parametrization is both most feasible and most useful in settings where $n_A << n_B$. In our application, the typical block has $n_A \approx 150$ and $n_B \approx 50,000$.

The second strategy deals with the common name problem without allowing flexibility at the record level. The affected fields in the comparison data can be modified to include a level indicating, rather than distance, that the record pair share a very common value. This is a computationally simple way to make records with less common names in blocks dominated by common names (think of, for example, men named Winston in a block dominated by men named William) more likely to be matched, and is suitable for situations where $n_A$ is large and record-specific parameters are impractical.

Both record-specific disagreement parameters and modification of the comparison data for very common names will be discussed in detail in Section \ref{method}.

\section{Record-specific Bayesian Method}
\label{method}

Our approach refines the $\beta$RL method of \cite{sad2017}. The assumption that the $\textbf{u}$ parameters are fixed is relaxed, allowing the parameters to vary across subgroups of records in $\textbf{A}$. In particular, these subgroups can be individual records, so that $\textbf{u}$ parameters can vary by record. Our implementation of the method is significantly faster than implementations of previous Bayesian methods, allowing for less aggressive blocking, record-specific parametrization and/or larger matching projects using typically available computing resources. Section \ref{imp} elaborates on our computational improvements. Code is available in the online supplement.

We first discuss construction of the comparison data from the original datafiles $\textbf{A}$ and $\textbf{B}$ in Section \ref{cd}, including our proposed lightweight method for modifying comparison data to include information on common names, which can be used with or without record-specific parameters. Section \ref{bm} sets out the model and discusses the refinement of the likelihood to allow record-specific $u_f$ parameters that vary by record $i$ in $\textbf{A}$. Section \ref{gs} and Section \ref{lf}, respectively, describe the Gibbs sampler and loss function we use.

\subsection{Comparison data}
\label{cd}

The comparison data is taken as given during the matching process, but must be constructed by the researcher from raw datafiles $\textbf{A}$ and $\textbf{B}$, which have $n_A$ and $n_B$ records, respectively, and where $n_A \leq n_B$, without loss of generality.

The raw datafiles have some shared number $F$ of variables, or fields, describing each record. $\textbf{A}$ and $\textbf{B}$ (for example, two censuses in different years) are thought to be measured from the same underlying population, or at least from two populations that significantly overlap.

To allow for flexibility across user-defined subgroups of records, a partition of the records in $\textbf{A}$ is defined with elements indexed by $g = 1, 2, \dots ,G$ for some $G \leq n_A$. Let $n_{A,g}$ be the number of records in the subgroup $g$. While these subgroups could be used to represent any partitioning of $\textbf{A}$ (see \cite{wortman2019} for an example of subgrouping using a small number of age ranges), their most useful application, and the application with which this paper is concerned, is when $G=n_A$, with each group $g$ corresponding to a record $i$ in $\textbf{A}$. Parameters that vary by record, which will be discussed below, are thus ``record-specific".

The comparison data itself, $\pmb{\gamma}(\textbf{A}, \textbf{B})$, contains comparisons $\gamma_{ij}^f$ for each pair of records $(i, j)$ from the two datafiles and for each field $f$. Comparison values are coded as positive integers, where $1$ indicates exact agreement and higher numbers indicate successively higher levels of disagreement up to the highest level for the field, $L_f$.

For string fields, we discretize the Jaro-Winkler string distance \citep[see][]{jaro1989, wink1990} between the two fields. The Jaro-Winkler distance is a standard string distance in record linkage, and roughly counts the number of edits needed to transform one string to the other, while weighting agreement in the first letter more heavily. Numeric fields such as birth year are compared by absolute difference, then discretized to the desired number of levels. Finally, for categorical fields such as state of birth, a binary measurement for exact agreement (1) or disagreement (2) is taken. 

Table \ref{tab:comp} gives the undiscretized and discretized comparison data for the toy datafiles in Table \ref{tab:simple}, where first name and last name distance are each discretized into four bins ($L_f=4$), and year of birth distance is discretized into four bins ($L_f=4$).

\begin{table}[ht]
\centering
\begin{tabular}{ll}
\begin{tabular}{c|c|c|c} 
(i,j)&First name & Last name & Year \\ \hline
(1,1)&0.00 & 0.163 & 0 \\ 
(1,2)&0.083 & 0.074 & 2 \\ 
(1,3)&0.542 & 0.562 & 3 \\ 
(1,4)&0.00 & 0.562 & 4 \\
(1,5)&0.542 & 1.00 & 4 \\
(2,1)&0.542 & 1.00 & 4 \\
(2,2)&0.514 & 1.00 & 6 \\
(2,3)&0.131 & 0.00 & 1 \\
(2,4)&0.542 & 0.00 & 0 \\
(2,5)&0.00 & 0.267 & 0 \\
\end{tabular}
\quad $\Longrightarrow$ \quad
\begin{tabular}{c|c|c|c} 
(i,j)&First name & Last name & Year \\ \hline
(1,1)&1 & 2 & 1 \\ 
(1,2)&2 & 2 & 2 \\ 
(1,3)&4 & 4 & 3 \\ 
(1,4)&1 & 4 & 3 \\
(1,5)&4 & 4 & 3 \\
(2,1)&4 & 4 & 3 \\
(2,2)&3 & 4 & 4 \\
(2,3)&2 & 1 & 2 \\
(2,4)&4 & 1 & 1 \\
(2,5)&1 & 3 & 1 \\
\end{tabular}
\end{tabular}
\caption{Undiscretized and discretized comparison data for $\textbf{A}$ and $\textbf{B}$ from Table \ref{tab:simple}. The first column indicates the record pair. Comparisons for first name and last name are Jaro-Winkler distances and comparisons for year of birth are absolute differences. }
\label{tab:comp}
\end{table}

\subsubsection{Adding information on common names}
\label{cdm}


In settings where record-specific parameters are not feasible, we propose including a disagreement level in the comparison data to indicate record pairs that have exact matches on common first names. This is done by constructing the comparison data as usual, then relabeling rows with exact matches on common names with the new disagreement level. Table \ref{tab:bincomp2} shows the modified comparison data when record pairs both with the first name ``John" are given their own disagreement level $\textbf{C}$.

\begin{table}[ht]
\centering
\begin{tabular}{c|c|c|c} 
(i,j)&First name & Last name & Year \\ \hline
(1,1)&\textbf{C} & 2 & 1 \\ 
(1,2)&2 & 2 & 2 \\ 
(1,3)&4 & 4 & 3 \\ 
(1,4)&\textbf{C} & 4 & 3 \\
(1,5)&4 & 4 & 3 \\
(2,1)&4 & 4 & 3 \\
(2,2)&3 & 4 & 4 \\
(2,3)&2 & 1 & 2 \\
(2,4)&4 & 1 & 1 \\
(2,5)&1 & 3 & 1 \\
\end{tabular}
\caption{Comparison data from toy problem in Table \ref{tab:simple} with added disagreement level $\textbf{C}$ to indicate that the record pair shares a common first name.}
\label{tab:bincomp2}
\end{table}

This procedure adds flexibility to the matching procedure but does not change the structure of the model or require computational modifications. While we code regular disagreement levels with integers for clarity, the label has no meaning in the model.

To see why adding a disagreement level for very common names improves matching, consider $\mathbf{m}_{f} =  (m_{f1}, \dots,  m_{fL_f})$, the vector of conditional probabilities that a matched record pair should have disagreement level $l$ in field $f$, and $\mathbf{u}_{f}=  (u_{f1}, \dots,  u_{fL_f})$, the equivalent for unmatched pairs.

Very common first names make $u_{f1}$, the probability of exact agreement ($l=1$) in field $f$ conditional on unmatched status, very large. This makes the likelihood ratio $m_{f1}/u_{f1}$ small, preventing accurate scoring of record pairs that should be categorized as matches. Tables \ref{tab:typblock} and \ref{tab:vcfn} give typical values for $\mathbf{m}_f$ and $\mathbf{u}_f$ in a typical and common first name block, respectively. Table \ref{tab:after} shows the effect of adding a disagreement level for very common first names (such as John or William)--a high likelihood ratio in the $l=1$ column is restored.

\begin{table}[ht]
\centering
\parbox{.45\linewidth}{
\centering
\begin{tabular}{c|c|c|c|c}
$l$ & 1 & 2 & 3 & 4 \\ \hline
$m_{fl}$ & 0.78 & 0.15 & 0.06 & 0.01 \\
$u_{fl}$ & 0.02 & 0.03 & 0.20 & 0.76 \\ \hline
$m_{fl}/u_{fl}$ & \textbf{39} & 5 & 0.3 & 0.013 \\
\end{tabular}
\caption{Selected parameter values in a typical block (illustrative, not real data).}
\label{tab:typblock}
}
\quad
\parbox{.45\linewidth}{
\centering
\begin{tabular}{c|c|c|c|c}
$l$ & 1 & 2 & 3 & 4 \\ \hline
$m_{fl}$ & 0.80 & 0.13 & 0.05 & 0.02 \\
$u_{fl}$ & \textbf{0.42} & 0.05 & 0.10 & 0.43 \\\hline
$m_{fl}/u_{fl}$ & \textbf{1.9} & 2.6 & 0.5 & 0.047 \\
\end{tabular}
\caption{Selected parameter values in a block with many common first names.}
\label{tab:vcfn}
}
\end{table}

\begin{table}[ht]
\centering
\begin{tabular}{c|c|c|c|c|c}
$l$ & Common & 1 & 2 & 3 & 4 \\ \hline
$m_{fl}$ & 0.40 & 0.40 & 0.13 & 0.05 & 0.02 \\
$u_{fl}$ & 0.40 & \textbf{0.02} & 0.05 & 0.10 & 0.43 \\\hline
$m_{fl}/u_{fl}$ & 1 & \textbf{20} & 2.6 & 0.5 & 0.047 \\
\end{tabular}
\caption{Selected parameter values when the comparison corresponding to Table \ref{tab:vcfn} includes a disagreement level for common names.}
\label{tab:after}
\end{table}

The effect of changing the comparison data in this way is not so much to improve match rates for records with very common names, which are difficult to match in any case, but to yield more matches among records with uncommon names located in a block dominated by very common names. Returning to the toy problem, we have a better chance of matching Jedediah Smith in $\textbf{A}$ if the large number of men named John do not dominate the $u_f$ distribution with spurious exact matches.

The choice of names (or other field values) to code as ``very common" for the purpose of modifying comparison data is the researcher's, but we recommend the number of such names be small so records with those names are not a large majority. However many records in a block are coded as having very common first names, an even higher proportion of record \emph{pairs} with exact agreement on first name will be marked as very common, because the number of such pairs rises in the product of the number of such records in each dataset. This can lead to problems identifying the probability of matching conditional on non-spurious exact agreements, because of a lack of data. In typical populations, there is little reason to mark more than the ten most common first names as very common.

\subsection{Model}
\label{bm}

In comparison to $\beta$RL, our model on the comparison data $\pmb{\gamma}$ is refined to allow for group- and record-specific disagreement probability parameters. We allow the prior on the matching to reflect the subgroup structure when subgroups are large enough to allow such parameters to be estimated. For simplicity we assume there are no missing data, but missingness can be easily accommodated under a missing-at-random assumption, as in \cite{sad2017}.

To prevent confusion between blocks and subgroups, it is noted here that a model is estimated individually for each block. In the discussion below, we abuse notation and use $\textbf{A}$ and $\textbf{B}$ to denote blocked subsets of the original datafiles rather than using more specific labels. On the other hand, subgrouping schemes to be discussed define subsets of the data \emph{within} blocks (record-specific groups being the principal example).

\subsubsection{Parameter of interest: the matching}
\label{poi}

It is possible to model latent data (i.e. the unobserved common population from which $\textbf{A}$ and $\textbf{B}$ are thought to be drawn), as in, for example, \cite{tancredi2011} and \cite{ste2015}, but it is more straightforward to model the matching itself. A matching labeling \citep{sad2017}, denoted $\mathbf{Z}$, is used to encode the relationship between the two observed datasets. For $\textbf{A}$ and $\textbf{B}$ with $n_A \le n_B$ records respectively, $\textbf{Z} = (Z_1, Z_2, \dots, Z_{n_A})$ is defined,
$$Z_i = \left\{
	\begin{array}{ll}
	j, & \text{if } i \text{ matches record } j \text{ in } \textbf{B}, \\
	0, & \text{if } i \text{ has no match in } \textbf{B}, \\
	\end{array}
	\right\}$$
where $i$ indexes records in $\textbf{A}$. A matching labelling encodes the same information as a matching matrix (an $n_A$ by $n_B$ matrix where 1 encodes a match and 0 a non-match), but is computationally easier to work with because of its smaller dimension.

\cite{sad2017} encoded a non-match, for record $i$, as $n_B+i$. The utility of this encoding is to allow the final step of obtaining a point estimate to be viewed as a linear sum assignment problem (LSAP). We set all non-matches to 0 for simplicity and because it reduces computational cost, but switching between encodings is trivial in cases where solving an LSAP is necessary to produce a point estimate (see Section \ref{lf}).

We let $n(\textbf{Z}) = \sum_{i=1}^{n_A} \textbf{1}(Z_i>0)$. Also, $n_{g}(\textbf{Z}) = \sum_{i=1}^{n_A} \textbf{1}(Z_i>0, i \in i_g)$ is the number of records in subgroup $g$ that have a match in $\textbf{Z}$. In addition to $\textbf{Z}$, the model includes the nuisance parameter $\mathbf{\Phi}$ containing the disagreement probabilities, defined below.

\subsubsection{Likelihood}
\label{ll}

The comparison data is modelled, conditional on a matching $\textbf{Z}$ and disagreement probabilities $\mathbf{\Phi}$, using the latent class framework introduced earlier. The likelihood is a modified version of the likelihood used in the $\beta$RL method of \cite{sad2017},
\[ \mathcal{L}(\mathbf{Z}, \mathbf{\Phi} | \pmb{\gamma}) = \prod_{g=1}^G \prod_{i: A_i \in g} \prod_{j=1}^{n_B} \prod_{f=l}^F \prod_{l=1}^{L_f} \left(m_{gfl}^{\textbf{1}(Z_i=j)} u_{gfl}^{\textbf{1}(Z_i \neq j)} \right)^{\textbf{1}(\gamma_{ij}^{f} = l)}, \]
where $\mathbf{Z}$ is the matching, $g$ indexes user-defined subgroups, $i$ indexes records in $\textbf{A}$ and $j$ indexes records in $\textbf{B}$, $\mathbf{\Phi} = (\mathbf{\Phi}_1, \dots, \mathbf{\Phi}_G)$, $\mathbf{\Phi}_g =  (\mathbf{m_g}, \mathbf{u_g}) = ( [\mathbf{m}_{g1}, \dots, \mathbf{m}_{gF}]^\top,  [\mathbf{u}_{g1}, \dots, \mathbf{u}_{gF}]^\top ) $ is the vector of disagreement probability parameters conditional on match status and $\pmb{\gamma}$ is the comparison vector between \textbf{A} and \textbf{B}. The vector $\mathbf{m}_{gf}$, corresponding to field $f$ for group $g$, is a vector of probabilities $(m_{gf1}, \dots, m_{gfL_f})$ conditional on $Z_i>0$ (matched) where $m_{gfl} = \text{P}(\gamma_{ij}^{f} =l | Z_i = j, A_i \in g)$. $\mathbf{u}_{gf}$ is defined similarly for unmatched status. The likelihood in \cite{sad2017} corresponds to setting $G=1$.

If the researcher chooses to take each record $i$ in $\textbf{A}$ as its own subgroup, the likelihood is
$$ \mathcal{L}(\mathbf{Z}, \mathbf{\Phi} | \pmb{\gamma}) = \prod_{i=1}^{n_A} \prod_{j=1}^{n_B} \prod_{f=l}^F \prod_{l=1}^{L_f} \left(m_{fl}^{\textbf{1}(Z_i=j)} u_{ifl}^{\textbf{1}(Z_i \neq j)} \right)^{\textbf{1}(\gamma_{ij}^{f} = l)}, $$
where the $\textbf{u}$ parameters are record-specific (note the $i$ subscript) and the $\textbf{m}$ parameters are fixed across records. Fixing the $\textbf{m}$ parameters is necessary for identification when each record $i$ is its own subgroup because there is so little information on pairs with matched status: there exists at most one match for each group/record from which to infer a distribution. This identifiability problem for the $\textbf{m}$ parameters can also be a concern with general subgroups $g$, if the groups are not sufficiently large. On the other hand, $\textbf{u}$ parameters can be modelled record-specifically, provided $n_B$ is large, because there are at least $n_B-1$ record pairs with non-match status in the group.



The appeal of record-specific $\mathbf{u}$ parameters is that the assumption of a fixed disagreement level distribution across records is relaxed and comparison data can be modelled differently across records in $\textbf{A}$. In the application to Union Army data, discussed in Section \ref{ua}, this refinement greatly improves performance in two ways: first, by achieving convergence in blocks where matches could previously not be identified and, second, by recovering more true matches in all blocks, particularly among men with less common names. To see why, consider Figure \ref{recwisetab}. \begin{figure}[ht]
\centering
\begin{tabular}{c|c|c|c|c}
\multicolumn{5}{c}{$u_{if} = u_f$}\\
$l$ & 1 & 2 & 3 & 4 \\ \hline
$m_f$ & \textbf{.80} & 0.14 & 0.05 & 0.01 \\ \hline
$u_f$ & \textbf{.03} & 0.05 & 0.19 & 0.73 \\
\end{tabular}
\quad \quad
\begin{tabular}{c|c|c|c|c}
\multicolumn{5}{c}{$u_{if}$ varies with $i$}\\
$l$ & 1 & 2 & 3 & 4 \\ \hline
$m_f$ & \textbf{.80} & 0.14 & 0.05 & 0.01 \\ \hline
$u_{1f} \text{ (common)}$ & \textbf{.10} & 0.12 & 0.26 & 0.52 \\
$u_{2f} \text{ (less common)}$ & \textbf{.01} & 0.08 & 0.20 & 0.71 \\
$u_{3f} \text{ (rare)}$ & \textbf{.003} & 0.067 & 0.20 & 0.73 \\
\end{tabular}
\caption{Typical disagreement probability parameters, with and without record-specific $u_{if}$. Field $f$ is last name, and the records 1, 2 and 3 have values ``Smith" (a common name), ``Stewart" (a less common name) and ``Stringham" (a rare name) respectively.}
\label{recwisetab}

\end{figure}This gives parameters for a matching of a hypothetical block of records with last names starting with $S$ and $Z$. The three records in $\mathbf{A}$ have values ``Smith" (a very common name), ``Stewart" (less common) and ``Stringham" (rare). In the table on the left, $u_{if}$ is fixed across records, while it varies in the table on the right. The highlighted column shows $u_{if1}$, the probability that the last name field has exact agreement in a record pair involving record $i$, conditional on the pair not being a match.

In the table on the right, record $3$ is associated with a likelihood ratio $m_f / u_{3f} = .80/.003 = 266.7$ for the last name field, a high likelihood ratio reflecting the fact that exact agreement on the name ``Stringham" is unlikely conditional on non-match status. Record $1$ is associated with a much lower likelihood ratio of $m_f / u_{1f} = .80/.10= 8$, since the name ``Smith" is much more common. In the table on the left, however, all likelihood ratios are constrained to be equal, since there is only one $u_{f}$ parameter across all records. Record $3$ in this case shows a much lower likelihood ratio of $26.7$, and will less likely be matched as a result.

In this way, our record-specific model most improves matching for records with less common names and other field values, while preventing false matches of records with very common names. Even with record-specific parameters, however, estimating within blocks invites the problem discussed by \cite{murray2016} of disagreement models that are distorted by the very process of blocking. ``Smith" might make up $10\%$ of names within a block of records with last names starting with S or Z, but only $2\%$ of names overall. If the distribution of disagreements over all record pairs across all blocks were used, the likelihood ratio would be even higher than in the above example and men named Smith would be more likely to be matched. The process of ``U-correction" in \cite{mcv20183} makes exactly this change, improving matching for records with common names. In our application in Section \ref{ua}, we estimate matchings with U-correction, both with and without record-specific parameters. We find that U-correction improves on the baseline model, but the improvement is less drastic than that from using record-specific parameters. When both procedures are combined, results are mixed but generally less successful than our preferred record-specific specification.

\subsubsection{Prior}
\label{prior}
The prior on the matching is the Beta prior introduced in \cite{sad2017}, but can be modified to accommodate subgroups,
$$
\nonumber
\text{P}(\textbf{Z}, \mathbf{p} | \alpha_{p}, \beta_{p})  = \frac{(n_B - n(\textbf{Z}))!}{n_B!} \frac{1}{\text{B}(\alpha_{p}, \beta_{p})^G}
 \prod_{g=1}^G \left [ p_g^{n_g(\textbf{Z})+\alpha_{p}-1} \left (1 - p_g\right)^{n_{A,g} - n_g(\textbf{Z})+ \beta_{p}-1} \right ],
$$
where $\text{B}$ is the Beta function, $\mathbf{p} = (p_1, p_2, \dots , p_G)$ and $p_g$ is an ex-ante matching probability for group $g$.
Allowing for subgroup-specific matching probabilities in the prior is only feasible when subgroups are sufficiently large. The posterior on $p_g$, conditional on $\textbf{Z}$, is a Beta distribution with parameters $n_g(\textbf{Z}) + \alpha_p$ and $\alpha_p, n_{A,g} - n_g(\textbf{Z}) + \beta_p$. We set $\alpha_p = \beta_p=1$, so $n_{A,g} \geq 50$ is likely sufficient to dominate the prior.


In the most important application of subgroups, where $G=n_A$ and each group $g$ corresponds to a record $i$ in $n_A$, record-specific $p_g$ is not feasible and the prior is simply that of \cite{sad2017}. The disagreement probabilities $\mathbf{m}_{gf}$ and $\mathbf{u}_{gf}$ for field $f$ have Dirichlet priors with hyperparameters $\bm{\alpha}_f=(\alpha_{f1}, \dots, \alpha_{fL_f})$.

\subsubsection{Posterior}

In the case where we use record-specific parameters, the posterior is
$$  \text{P}(\textbf{Z}, \mathbf{\Phi}, \mathbf{p} | \pmb{\gamma}, \alpha_p, \beta_p, \{\bm{\alpha}_f\}_{f=1}^F) \propto \text{P}(\textbf{Z},\mathbf{p} | \alpha_{p}, \beta_{p}) \frac{1}{\text{B}(\bm{\alpha}_f)} \prod_{f=1}^F \prod_{l=1}^{L_f} m_{fl}^{n_{m}^{fl} + \alpha_{fl}-1} \left[ \prod_{i=1}^{n_A}  u_{ifl}^{n_{u}^{ifl} + \alpha_{fl}-1 } \right], $$
where $\text{B}$ is the multivariate Beta function, $n_{m}^{fl} =  \sum_{i=1}^{n_A} \sum_{j=1}^{n_B} \mathbf{1}(Z_i=j, \gamma_{ij}^{f}=l) $, $n_{u}^{ifl} = \sum_{j=1}^{n_B} \mathbf{1}(Z_i \neq j, \gamma_{ij}^{f}=l) $ and the prior for $\textbf{Z}$ and $\textbf{p}$ is as specified above. In words, $n_{m}^{fl}$ is the number of record pairs which are matched in $\mathbf{Z}$ and for which field $f$ has disagreement level $l$, and $n_{u}^{ifl}$ is the number of unmatched pairs for record $i$ for which field $f$ has disagreement level $l$. This posterior is equivalent to the posterior in $\beta$RL if $\mathbf{p}$ is marginalized out, $u_{ifl}$ is set to $u_{fl}$ for all $i$ and $\sum_{i=1}^{n_A}n_{u}^{ifl}$ is replaced by $n_{u}^{fl}$.


We draw from the posterior using the Gibbs sampler presented in \cite{sad2017}, with modifications to allow subgroups and record-specific parameters, as explained below.

\subsection{Gibbs Sampler}
\label{gs}

The Gibbs sampler is initialized with an empty matching $\mathbf{Z}$ such that $Z_i=0$ for all $i$. Then the following steps are repeated for a specified number of iterations:
\begin{enumerate}
	\item{The matching probability parameters $\mathbf{p}_g$ are drawn conditional on the most recent draw for $\mathbf{Z}$. For group $g$, $p_g$ is distributed Beta with parameters $n_g(\textbf{Z}) + \alpha_p$ and $n_{A,g} - n_g(\textbf{Z}) + \beta_p$. 
}
	\item{The disagreement probability parameters $\mathbf{\Phi}$ are drawn conditional on the most recent draw for $\mathbf{Z}$. For field $f$, $\mathbf{m}_{f} = (m_{f1}, \dots, m_{fL_f})$ is distributed Dirichlet with parameter $\mathbf{n}_{m}^{f} + \bm{\alpha}_f$ where $\mathbf{n}_{m}^{f} = (n_{m}^{f1}, \dots, n_{m}^{fL_f})$ and $\bm{\alpha}_f = (\alpha_{f1}, \dots, \alpha_{fL_f})$. For group $g$, $\mathbf{u}_f$ is distributed Dirichlet with parameter $\mathbf{n}_{u}^{gf} + \bm{\alpha}_f$, where $\mathbf{n}_{u}^{gf} = (n_{u}^{gf1}, \dots, n_{u}^{gfL_f})$. 
}
	\item{For each record $i$ in group $g$ in $\textbf{A}$, $Z_i$ is drawn conditional on the most recent draws for each record in $\mathbf{Z}_{-i}$, $\mathbf{p}_g$ and $\mathbf{\Phi}$.
Given $\textbf{Z}_{-i}$, where $\textbf{Z}_{-i}$ is all elements of $\textbf{Z}$ other than that for record $i$ in $\textbf{A}$, $Z_i$ has a discrete distribution,\begin{multline}\text{P}(Z_i =j | \textbf{Z}_{-i}, \mathbf{\Phi} , \mathbf{p}, \pmb{\gamma}, \alpha_p, \beta_p, \{\bm{\alpha}_f\}_{f=1}^F ) \propto \left(n_B - n(\textbf{Z}_{-i})\right) \textbf{1}(j=0) + \\ \frac{p_g}{1-p_g} \text{exp}\left(\ \sum_{f=1}^F \sum_{l=1}^{L_f} \textbf{1}(\gamma_{ij}=l) \text{log} \left(\frac{m_{fl}}{u_{gfl}} \right) \right)  \textbf{1}(j>0)\textbf{1}(Z_{i'} \neq j, \forall i' \neq i),
\label{step3}
\end{multline}
where $n(\mathbf{Z}_{-i}) = \sum_{i'=1}^{n_A} \textbf{1}(Z_{i'}>0, i' \neq i)$. The first summand gives the unnormalized probability of record $i$ drawing a non-match, and the second the probability of matching to any particular record $j$ in $\textbf{B}$. 
}
\end{enumerate}

In Equation (\ref{step3}), the logarithm term inside the second summand has a straightforward interpretation as a log-likelihood ratio. $m_{fl}$ gives the likelihood that a true match should have produced the observed disagreement level $l$ in field $f$, while $u_{fl}$ gives the likelihood that a true non-match should have produced it. 

We note that the term $\textbf{1}(j>0)\textbf{1}(Z_{i'} \neq j, \forall i' \neq i)$ in Equation (\ref{step3}) enforces a bipartite matching at each iteration: record $i$ cannot be matched to record $j$ if record $j$ is currently matched to some $i'$. This feature, while necessary, can lead to the sampler getting ``stuck" repeatedly matching record $i$ to $j$ when $j$ should be matched to a record $i'$ for which the match is drawn later. To avoid favoring records in $\mathbf{A}$ that happen to appear before others, rather than drawing the matching in a fixed order, we randomize the order in which records $i=1, \dots, n_A$ have matches drawn. This also allows the researcher to test for stuck matches by running multiple chains, each of which will draw matches in its own, random order.

Our implementation of this Gibbs sampler achieves a significant improvement in speed over the implementation in the {\tt BRL} R package developed for \cite{sad2017}---computation is discussed in Section \ref{imp}.

Alternative sampling methods were explored, including the novel local balancing method of \cite{zan2019}, designed for high-dimensional discrete spaces. We implemented a version of that method, but found that while it performed well given a sufficiently large number of burn-in steps, the computational expense of exploring the space of matchings was prohibitive in large blocks (such as in our application in Section \ref{ua}). This computational difficulty was also noted by \cite{mcv20183}.

\subsection{Loss function}
\label{lf}

The loss function of \cite{sad2017} is used to yield a point estimate. For expository convenience, we omit here the option to leave record pairs unassigned and the corresponding penalty for the unassignment decision. The loss function for a Bayes estimate $\hat{\mathbf{Z}}$ is $L(\mathbf{Z}, \hat{\mathbf{Z}}) = \sum_{i=1}^{n_A} L(Z_i, \hat{Z}_i)$, where $L(Z_i, \hat{Z}_i)$ is
\begin{equation}
\label{lfdef}
L(Z_i, \hat{Z}_i) = \left\{ 
	\begin{array}{ll}
	0, & \text{if } Z_i = \hat{Z}_i; \\
	\lambda_{FNM}, & \text{if } Z_i > 0, \hat{Z}_i = 0; \\
	\lambda_{FM1}, & \text{if } Z_i = 0, \hat{Z}_i > 0; \\
	\lambda_{FM2}, & \text{if } Z_i, \hat{Z}_i > 0, Z_i \neq \hat{Z}_i,
	\end{array}
	\right\}
\end{equation}
where $\hat{Z}_i$ indicates the estimate of $\textbf{Z}$ for record $i$ in $\textbf{A}$, $\lambda_{FNM}$ is the loss from assigning a non-match when a match should be assigned, $\lambda_{FM1}$ is the loss from assigning a match when there is none, and $\lambda_{FM2}$ is the loss from matching to the wrong element of $\mathbf{B}$. 

\begin{thm} \label{loss_func}
Given the loss function in Equation (\ref{lfdef}), if $0 < \lambda_{FNM} \leq \lambda_{FM1}$ and $\lambda_{FM2} \geq \frac{3 \lambda_{FNM} + \lambda_{FM1}}{2}$, the Bayes estimate $\hat{\mathbf{Z}}$ has elements
\[ \hat{Z}_i =  \left\{
	\begin{array}{ll}
	j, & \text{if } \text{P}(Z_i =j | \pmb{\gamma}) > \frac{\lambda_{FM1}}{\lambda_{FM1} + \lambda_{FNM}} + \text{  P}(Z_i \notin \{0, j\}| \pmb{\gamma}) \frac{\lambda_{FM2}-\lambda_{FM1} - \lambda_{FNM}}{\lambda_{FM1}+\lambda_{FNM}} \\
	0, & \text{otherwise,} \\
	\end{array}
	\right\}  \]
where probabilities are posterior probabilities.
\end{thm}

This theorem gives the same conclusion as Theorem 1 in \cite{sad2017} but under slightly weaker conditions. The proof, including a proof that the conditions are weaker, is in the appendix. In the application, $\lambda_{FNM} = \lambda_{FM1} = 1$ and $\lambda_{FM2}=2$ are taken as default values, though estimates at other values are also produced. Then, the estimate is given by
\[ \hat{Z}_i =  \left\{
	\begin{array}{ll}
	j, & \text{if } \text{P}(Z_i =j | \pmb{\gamma}) > 1/2 \\
	0, & \text{otherwise.} \\
	\end{array}
	\right\}  \]
For general values of the lambda parameters, the problem is a linear sum assignment problem (LSAP), a class of problems to which there are well-known solution methods. For very large problems, however, solving the LSAP involves considerable computational expense.


\subsection{Estimating TPR \& PPV} \label{tprppv}

When true match status is unavailable, TPR (true positive rate) and PPV (positive predictive value) can be estimated from the posterior Gibbs sample. Given an estimate $\hat{\textbf{Z}}$ and all matchings $\mathbf{Z_t}$ in the sample, where $t = 1, \dots, T$ indexes the iterations of the Gibbs sampler and $T$ is the number of iterations less the number dropped as burn-in, we have $\hat{TPR}(\hat{\textbf{Z}}) =  \frac{1}{T}\sum_{t=1}^T TPR(\hat{\textbf{Z}}, \mathbf{Z_t})$ and $\hat{PPV}(\hat{\textbf{Z}}) =  \frac{1}{T}\sum_{t=1}^T PPV(\hat{\textbf{Z}}, \mathbf{Z_t})$, where
\[
TPR(\hat{\textbf{Z}}, \mathbf{Z_t}) = \frac{ \sum_{i=1}^{n_A} \mathbf{1} (\hat{Z}_i = Z_{ti}, \hat{Z}_i>0) }{\sum_{i=1}^{n_A} \mathbf{1}(Z_{ti}>0) },
\]
and
\[
PPV(\hat{\textbf{Z}}, \mathbf{Z_t}) = \frac{ \sum_{i=1}^{n_A} \mathbf{1} (\hat{Z}_i = Z_{ti}, \hat{Z}_i>0) }{\sum_{i=1}^{n_A} \mathbf{1}(\hat{Z}_{i}>0) }.
\]
In words, TPR and PPV for $\hat{\textbf{Z}}$ are directly computed for each matching in the Gibbs sample, then averaged. These quantities converge to the posterior expectations of $TPR(\hat{\textbf{Z}}, \mathbf{Z})$ and $PPV(\hat{\textbf{Z}}, \mathbf{Z})$ for $\hat{\textbf{Z}}$.

\section{Application - Union Army}
\label{ua}

The Union Army dataset, created by the Early Indicators project \citep{fog00}, covers a sample of white, Northern men who served in the Union Army during the US Civil War, linking individual records across census, military and administrative data. Matches were made manually by experts, at considerable expense, so we can take them as ``true" matches for evaluating the output of automated matching procedures, despite the absence of unambiguous unique identifiers.

There are 39,354 men in the Union Army military dataset, of whom 15,485 have records suitably complete for automated matching. 8,628 of these are recorded as having high-quality matches, while 1,016 are matched with less certainty and 5,841 have no known match. We discard the 1,016 records with uncertain matches so that TPR and PPV can be meaningfully evaluated, leaving 14,469 records. This military dataset is dataset $\textbf{A}$ in the model, the smaller of the two datasets.

The second dataset $\textbf{B}$ is the digitized full-count 1900 US Census, obtained through the National Bureau of Economic Research (NBER). After filtering for white men born no later than 1850, there are 4,657,508 complete records. $\textbf{A} \times \textbf{B}$ thus contains over 67 billion record pairs. After blocking, there are about 754 million pairs (see Section \ref{appblocking}).

Reliable fields common to both datafiles are first name, last name, birth year and state/county of birth. In preparation for matching, names were split into first and last name fields and parenthetic comments, titles and middle initials were removed. Capitalization was standardized. Common abbreviated first names such as Wm., Geo., and Thos. were replaced with spelled-out equivalents (William, George and Thomas, respectively). Country and state names in the birthplace field were standardized to their equivalents in the Union Army data.

To construct comparisons for the name fields, we used Jaro-Winkler string distances, mentioned above. The distances were discretized into seven bins, with cutpoints at 0.05, 0.1, 0.15, 0.22, 0.3, and 0.45. Reducing the number of bins or moving the cutpoints does not significantly affect the estimated matching. Comparisons for birth year are absolute differences between years, discretized in four bins, with cutpoints at 1.5, 2.5 and 4.5. Birthplace comparisons are a binary variable indicating whether there is an exact match.

\subsection{Blocking}
\label{appblocking}

Data is not blocked on year or state of birth. We block on initials, grouping them based on frequency and common errors. Blocks for first name are, where letters indicate initials included in the block, AEIOUY, BFP, CKQSXZ, DT, GJ, Jo, HMN, LR, and VW. The Jo block includes men whose first name begins with Jo and was necessary because of the large number of Johns and Josephs. Blocks for last name are AEIOUY, B, CKQX, DT, FP, GJ, H, LNR, M, SZ, VW. This scheme yields 99 roughly equally-sized blocks which contain among them approximately 754 million pairs of records.

We note that computation can be straightforwardly parallelized across blocks, though this was not done for the present application as estimation was already sufficiently fast to be practical.

\subsection{Estimation}

The hyperparameters on the Beta distribution on the matching probabilities were set to $\alpha_p = \beta_p = 1$. The Dirichlet hyperparameters on the disagreement probabilities were set to $\{\{\alpha_{fl} \}_{l=1}^{L_f}\}_{f=1}^F =1$. Results are not sensitive to deviations from this assumption of flat priors. In the loss function, $\lambda_{FNM} = \lambda_{FM1} = 1$ and $\lambda_{FM2}=2$, as mentioned above.

Several different matchings on the Union Army data were estimated. The first was a baseline matching, equivalent to $\beta$RL. In the second, exact matches on the eight most common first names (John, William, James, George, Charles, Henry, Thomas and Joseph) were encoded as their own disagreement level within the first name field. As noted in Section \ref{cdm}, the number of very common names encoded this way should be small, typically no greater than ten for first names. The number of names we took as very common (eight) was arbitrary within this range. In the third specification, we allowed for record-specific $\textbf{u}$ parameters (see Section \ref{ll}). We also estimated a matching using both methods in combination.

Each Gibbs sampler was run for 1,000 iterations, with the first 100 iterations dropped as burn-in (results were virtually identical when we set the number of iterations to 2,500 in tests). Parallel chains were not run for the main match, because performance was already very good and there was consistent convergence as indicated by Gelman-Rubin statistics. 

Matching took 16 and 18 minutes using a single core on a standard Linux workstation, blocking as above, for the first and second specifications, respectively. Our record-specific specification took 61 minutes. The combined specification took 66 minutes. All timings include construction of comparison data, which in each case took about three minutes.

We also adapted the U-correction procedure of \cite{mcv20183} to our method and performed matchings for each of our specifications with and without this procedure, computing margins of disagreement levels across all record pairs for the blocked variables (first and last name). In the record-specific case, we computed margins for each record in $\mathbf{A}$. Computing these margins before running the main model took 21 minutes, but they were reused over specifications.

In comparison, matching using the R package {\tt BRL} of \cite{sad2017}, including comparison data construction, would take roughly 12 hours, based on our extrapolation of timing from a sample of blocks. We also estimated the matching using the {\tt BayesianRecordLinkage} Julia package of \cite{mcv20183}, which also uses the Bayesian model of \cite{sad2017}. We used the same initial blocking procedure as in our matching. We found that the post hoc blocking procedure struggled with the asymmetry in the size of datafiles, so that we needed to raise the maximum post hoc block size from the recommended 2,500 to 12,800. We ran the MCMC procedure for 10,000 steps. Estimation took approximately 8.5 hours including construction of comparison data. No matches were returned at reasonable loss function parameters, likely meaning more steps are needed to achieve convergence.

For comparison, matchings using the frequentist method of \cite{imai2019} as implemented in their package {\tt fastLink} were also run, both with and without the option to reweight parameters on first name comparisons by frequency. We used the same blocking scheme as with our method. First and last name were allowed to match partially. Recommended parameters were used. Computation times were 22 minutes without reweighting and 24 minutes when reweighting was done. Computations were run in parallel on three cores and timings include construction of comparison data. 

The option to reweight parameters by first name frequency in {\tt fastLink} is given as a way to deal with the common name problem, at least for first names, and can be compared to our methods of modifying comparison data and allowing record-specific disagreement parameters. We also investigated using the ``soft" TF-IDF string distance metric of \cite{cohen2003}, using a recent implementation, as an alternative approach to dealing with the common name problem, but computational costs were prohibitive in our setting, so we have left a comparison to future work.

We estimated a match using the EM method of \cite{abr2019}, implementing the method in R. We attempted a matching using our blocking scheme, but this was computationally infeasible, at least with our implementation. Using the blocking scheme of \cite{abr2019}, which blocks on first and last initials, state of birth, and year of birth within a five year window, computation time was 20 minutes, including construction of comparison vectors. We attempted an intermediate blocking scheme using letter groups rather than initials (while still blocking on state and year of birth) but no correct matches were found in this case.

\begin{figure}[ht]
\includegraphics[scale=0.9]{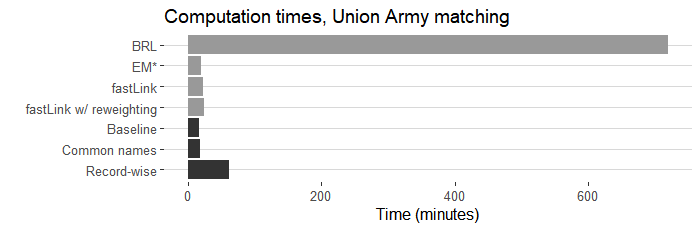}
\caption{All computations in R. Darkened bars indicate our new method. *The EM matching was estimated with a finer blocking scheme than the other methods.}
\label{osp}
\end{figure}

Overall timings for the methods mentioned are shown in Figure \ref{osp}. The method of \cite{mcv20183}, discussed above, is omitted as convergence was not achieved. We note that our method significantly improves on {\tt BRL} in computing time. Our implementation is competitive with frequentist methods, including the high-performance {\tt fastLink} package, even when record-wise parameters are used. Computation speed is further discussed in section \ref{imp}.

\subsection{Results}

\begin{figure}[ht]
\includegraphics[scale=0.68]{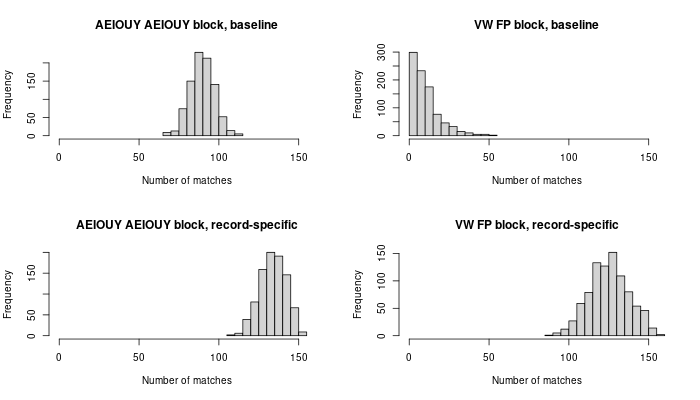}
\caption{Posterior distributions of the number of matches in the AEIOUY AEIOUY and VW FP blocks, for the baseline (top row) and record-specific (bottom row) specifications.}
\label{histplot}
\end{figure}

Our procedure returns a posterior distribution on the $\mathbf{p}$, $\mathbf{m}$ and $\mathbf{u}$ parameters, as well as on the matching itself, within each block. Figure \ref{histplot} shows the posterior distribution of the number of true matches in the baseline and record-wise specifications in two different blocks, AEIOUY AEIOUY and VW FP (refer to Section \ref{appblocking} for details about the blocking scheme). The AEIOUY AEIOUY block is a typical block in which the baseline model converged, while the VW FP block is made up mainly of men with the first name William, and in which the baseline model struggled to make links. Our record-specific method estimates more matches than the baseline in both blocks, but the difference is starker in the VW FP block, where the mode for the number of matches in the baseline model is zero. The record-specific method overcomes the problem of spurious exact agreement in blocks dominated by common names discussed in Section \ref{ll}.

\subsubsection{Precision and recall}

Because expert-linked matches were available, true positive rate (TPR) and positive predictive value (PPV) can be computed. TPR, also called recall, is the number of true matches identified by the algorithm, divided by the total number of true matches. PPV, also called precision, is the number of true matches identified by the algorithm, divided by the number of matches identified by the algorithm.

\begin{figure}[ht]
\includegraphics[scale=0.86]{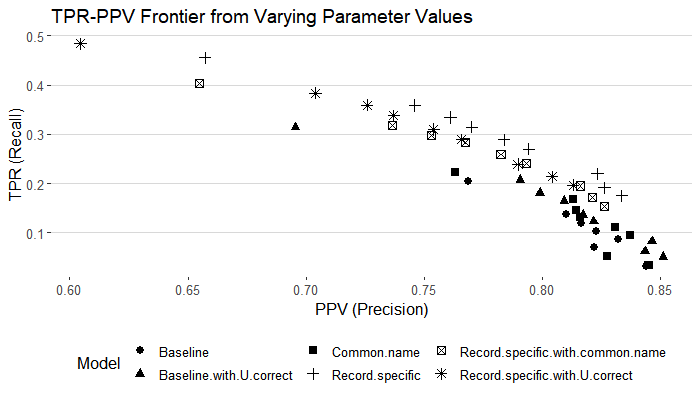}
\caption{Actual TPR and PPV values for selected matchings. Moving in a northwest direction toward higher PPV and lower TPR, along the frontiers, corresponds to increasing $\lambda_{FM2}$ and $\lambda_{FM1}$ relative to $\lambda_{FNM}$ while $\lambda_{FM2}/\lambda_{FM1}$ is held equal to 2.}
\label{frontier}
\end{figure}

With $\lambda_{FNM} = \lambda_{FM1} = 1$ and $\lambda_{FM2}=2$, the baseline specification of the Bayesian method yielded a TPR of 0.204 and a PPV of 0.768. Augmenting the comparison data with a level for very common first names gave a modestly improved TPR of 0.223 and PPV of 0.763. Our preferred approach, using record-specific $\mathbf{u}$ parameters, achieved a TPR of 0.455 and a PPV of 0.659. Higher accuracy is achieved by more heavily penalizing false matches in the loss function, so that our record-specific model easily dominates the other two along the TPR/PPV frontier, yielding 50-100\% more true matches when accuracy is held constant (see Figure \ref{frontier}). 

\begin{figure}[ht]
\includegraphics[scale=0.86]{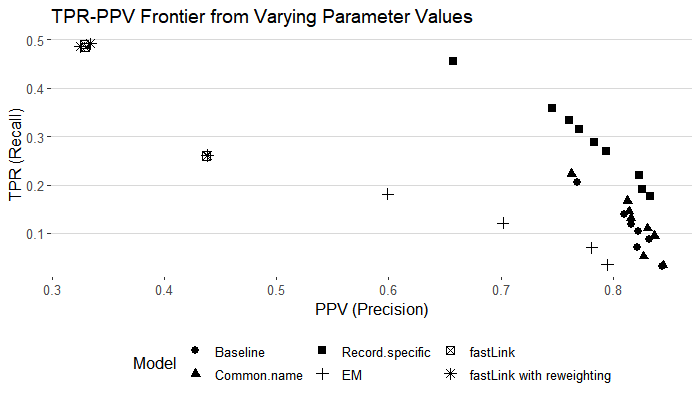}
\caption{Actual TPR and PPV values for selected matchings. For our specifications, moving in a northwest direction toward higher PPV and lower TPR, along the frontiers, corresponds to increasing $\lambda_{FM2}$ and $\lambda_{FM1}$ relative to $\lambda_{FNM}$ while $\lambda_{FM2}/\lambda_{FM1}$ is held equal to 2.}
\label{frontier2}
\end{figure}

The U-correction procedure noticeably improved performance over the baseline specification, but fell short of the record-specific specification. When both methods were combined by using record-specific parameters and a record-specific U-correction procedure, results were slightly worse than using record-specific parameters alone. Figure \ref{frontier} plots values of TPR and PPV for all of our specifications across different values of loss function parameters.
Figure \ref{frontier2} gives results for EM and {\tt fastLink} methods, along with our baseline and record-specific specifications.


The above measures of TPR and PPV are ``true" in the sense that they are drawn from expert-linked matches available for the Union Army dataset. TPR and PPV can also be estimated within the model when true matches are unavailable, as discussed in Section \ref{tprppv}. Estimated rates for our specifications are listed in Table \ref{pttable}. Estimated TPR and PPV were available for {\tt fastLink} matchings but we found they were unreliable in settings where $n_A << n_B$.
\begin{table}[ht]
\centering
\begin{tabular}{r|r|r|l|l}
Method & Specification & Parameters & TPR (est. TPR)& PPV (est. PPV) \\
\hline
Bayesian &baseline & $\lambda = (1,1,2)$ & 0.204 (0.326) &0.768 (0.756) \\
-&common names & $\lambda = (1,1,2)$ &0.223 (0.361) &0.763 (0.773)\\
-&record-specific &$\lambda = (1,1,2)$ &0.455 (0.389) &0.657 (0.824)\\
-&record-specific &$\lambda = (1,2,4)$ &0.359 (0.301) &0.746 (0.922) \\
-&record-specific &$\lambda = (1,3,6)$ &0.315 (0.263) &0.770 (0.949)\\

\end{tabular}
\caption{Actual against estimated TPR and PPV for selected specifications.}
\label{pttable}
\end{table}

Discrepancies between error rates as specified in the model and actual error rates are a common feature of record linkage models; see \cite{bel1995} and \cite{lar2001} for discussions. One possible explanation for the discrepancies here is a violation of the conditional independence assumption. Using true match status, there is a 7\% correlation between exact agreement on first name and exact agreement on last name, conditional on matching. In future work, this dependence could be integrated in a log-linear model of the comparison data by allowing for interaction between first and last name \cite[see][]{thib1993}, though \cite{xu2019} argue this technique does not reliably yield better matchings.

The relatively low TPR and PPV seen here is typical of automated methods, which cannot encode all information that human linkers can (here we are limited to four variables). The size asymmetry ($n_A < n_B$) problem and small $n_A$ problem, discussed in Section \ref{prac}, also make matching significantly more difficult. However, the TPR and PPV values found for all methods are likely underestimates, as some of the records in $\mathbf{A}$ with no expert-linked matches may in fact have matches in the data.

\section{Implementation}
\label{imp}

We implemented the method in R \citep{Rcite}, with the core segment of the Gibbs sampler written in C++, using the {\tt Rcpp} package for R \citep{Rcppcite} as an interface.


This implementation is faster than the latest available version of the {\tt BRL} package when using the same specification, achieving an order of magnitude improvement in computation time. This advantage is even greater when $n_A << n_B$ (see plots below). Even our record-specific method, which is considerably more complex, runs more quickly than {\tt BRL}.

Some speed is gained, relative to {\tt BRL}, by coding non-matches as 0 instead of using unique codes for each record in $\textbf{A}$, and by taking $\textbf{A}$ to be the smaller dataset rather than the larger. Most of the gains, however, come from purely computational improvements. When $n_A < n_B/200$, we take advantage of the sparsity of matches (at most $n_A$ of the $n_A n_B$ potential pairs are matches because of the bipartite restriction) by pre-computing likelihood sums in the $\textbf{Z}$-sampling step. We compute the total likelihood over records in $\mathbf{B}$ for each record in $\mathbf{A}$ using matrix multiplication, then subtract the contributions of the few elements in $\mathbf{B}$ already assigned to a match, rather than adding up all needed likelihood contributions in the main loop in Step 3 of the sampler.

Even without pre-computing likelihood sums, our implementation runs much more quickly when $n_A << n_B$, holding $n_A n_B$ fixed. We use vectorized functions and matrix multiplication over records in $\mathbf{B}$ whenever possible, rather than loops. This is generally not possible over records in $\mathbf{A}$, which are sampled sequentially. The speed advantage from having $n_A << n_B$ is even more significant when record-specific parameters are used, as the number of record-specific $\mathbf{u}$ parameters is proportional to $n_A$. 

We also aggressively minimized memory allocation and de-allocation in our code. Comparison data are constructed only for unique pairs of field values (for example, the comparison between the names ``John" and ``Joseph" is made only once), and the comparison vectors across fields are directly constructed in hashed form. Data are only stored unhashed for unique values of the final comparison vectors, of which there are far fewer than $n_A n_B$. Whenever possible, we use tabulations of comparison vector values over records, rather than the comparison data themselves. For example, when computing tabulations of record pairs by disagreement levels, we first tabulate record pairs by unique hashed comparison vector values, then matrix multiply by the unhashed binary matrix that records disagreement levels for each unique value, rather than directly tabulating over each disagreement level.

\begin{figure}[ht]
\centering
\includegraphics[scale=0.9]{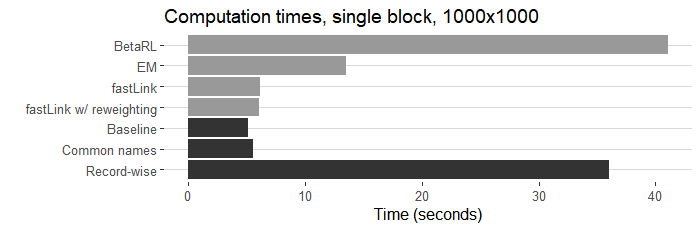}
\caption{Average computation time over several trials when $n_A=n_B=1000$. Computations in R. Darkened bars indicate our methods. All methods use default parameters.}
\label{speedplot}
\end{figure}
\begin{figure}[ht]
\centering
\includegraphics[scale=0.9]{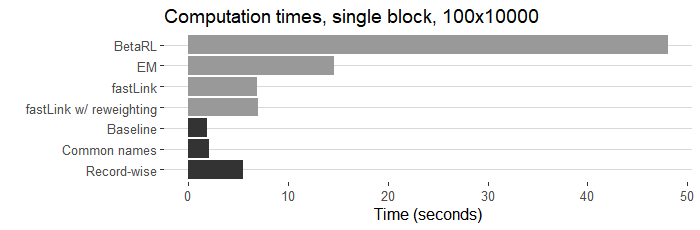}
\caption{Average computation time over several trials when $n_A=100$, $n_B=10000$. Computations in R. Darkened bars indicate our methods. All methods use default parameters.}
\label{speedplot2}
\end{figure}

To illustrate the additional speed gains from our implementation in the case where $n_A$ is much smaller than $n_B$, Figure \ref{speedplot} gives computation times for a 1000 by 1000 single-block matching problem, while Figure \ref{speedplot2} gives the computation times for a 100 by 10000 matching problem. 
Both problems are subsets of the Union Army problem. In both cases, our baseline method is faster than {\tt BRL}, to which it is equivalent in specification, and {\tt fastLink}, a recent frequentist method. Our record-specific method is significantly slower than our baseline method because of the significantly greater number of parameters, but the speed is still comparable to other methods, particularly when the ratio $n_B/n_A$ is high.

We note that these relative timings cannot necessarily be scaled up by the number of record pairs to give estimates of timings for larger blocks. For example, parallelization in {\tt fastLink} imposes fixed costs that make matchings relatively faster in bigger problems (see Figure \ref{osp}).

\section{Discussion} \label{discussion}

Existing Bayesian record linkage methods can perform poorly in large applications when blocks are dominated by common names or when $n_A << n_B$. Models of comparison data can be misspecified when there is spurious exact agreement between record pairs with very common field values, and asymmetry in datafile sizes can lead to large numbers of false positives. At the same time, Bayesian methods are computationally demanding, leading to a steep tradeoff between speed and flexibility.

Our proposed method and implementation push through that tradeoff by speeding up computation by an order of magnitude. Parallel computing methods could be used to reduce computation time further. Our more flexible Bayesian model allows for record-specific disagreement parameters and can yield significantly better results along the precision/recall frontier, particularly in large matching problems where many records have common field values. However, it is still limited by the need for manual blocking; an ``end-to-end" procedure that incorporates blocking into the model, in the spirit of \cite{marchant2021}, is a promising path for future research.




\newpage
\appendix
\section*{Appendix: Proof of Theorem \ref{loss_func}}
\begin{proof}
The proof is similar to the proof of Theorem 1 in \cite{sad2017}. The Bayes estimate $\hat{\mathbf{Z}}$ is the minimizer of the posterior expected loss, which is written $$ \mathbb{E}(L(\mathbf{Z}, \hat{\mathbf{Z}})|\pmb{\gamma}) = \sum_{i=1}^{n_A} \mathbb{E}(L(Z_i, \hat{Z}_i)|\pmb{\gamma}),$$
where \begin{equation}
\mathbb{E}(L(Z_i, \hat{Z}_i)|\pmb{\gamma}) = \left\{
	\begin{array}{ll}
		\lambda_{FNM} \text{P}(Z_i \neq 0 | \pmb{\gamma}), & \text{if } \hat{Z}_i = 0; \\
		\lambda_{FM1} \text{P}(Z_i = 0 | \pmb{\gamma}) + \lambda_{FM2} \text{P}(Z_i \notin \{0, j\} | \pmb{\gamma}), & \text{if } \hat{Z}_i = j >0.
	\end{array}
	\right.
\end{equation}
Minimizing $\mathbb{E}(L(Z_i, \hat{Z}_i)|\pmb{\gamma})$ for each $i$, ignoring other records, yields the theorem rule,
\begin{equation}
\label{minim}
 \hat{Z}_i =  \left\{
	\begin{array}{ll}
	j, & \text{if } \text{P}(Z_i =j | \pmb{\gamma}) > \frac{\lambda_{FM1}}{\lambda_{FM1} + \lambda_{FNM}} + \text{  P}(Z_i \notin \{0, j\}| \pmb{\gamma}) \frac{\lambda_{FM2}-\lambda_{FM1} - \lambda_{FNM}}{\lambda_{FM1}+\lambda_{FNM}} \\
	0, & \text{otherwise,} \\
	\end{array}
	\right\}
\end{equation}
as can be easily seen by rearranging the inequality $\mathbb{E}(L(Z_i, j^*)|\pmb{\gamma}) > \mathbb{E}(L(Z_i, 0)|\pmb{\gamma})$, where $j^*$ is the record $j$ for which $P(Z_i=j)$ is greatest (and thus the potential loss from picking it the smallest, since loss parameters are strictly positive).

This marginal solution minimizes the posterior expected loss if it produces a bipartite matching, where no two records in $\textbf{A}$ are matched to the same record in $\textbf{B}$ and vice versa. A sufficient condition for this is if the right hand side of the first condition in (\ref{minim}) is at least $1/2$, so that $\text{P}(Z_i=j|\pmb{\gamma})$ is greater than $\text{P}(Z_{i'}=j|\pmb{\gamma})$ for any other $i'$ (recall the posterior of $\mathbf{Z}$ is a distribution over bipartite matchings). If the theorem conditions $0 < \lambda_{FNM} \leq \lambda_{FM1}$ and $\lambda_{FM2} \geq \frac{3 \lambda_{FNM} + \lambda_{FM1}}{2}$ hold,
\begin{align*}
\overset{\lambda_{FM2} \geq \frac{3 \lambda_{FNM} + \lambda_{FM1}}{2}}{\implies} \frac{1}{2} (\lambda_{FM1} - \lambda_{FNM}) + \lambda_{FM2} - \lambda_{FM1} - \lambda_{NM}& \geq 0 \\
\overset{\lambda_{FNM} + \lambda_{FM1}>0}{\implies} \frac{\lambda_{FM1} - \lambda_{FNM}}{2(\lambda_{FM1} +\lambda_{FNM})} + \frac{\lambda_{FM2} - \lambda_{FM1} - \lambda_{NM}}{\lambda_{FM1} +\lambda_{FNM}}& \geq 0 \\
\overset{\lambda_{FM1}-\lambda_{FNM} \geq 0}{\implies} \frac{\lambda_{FM1} - \lambda_{FNM}}{2(\lambda_{FM1} +\lambda_{FNM})} + p\frac{\lambda_{FM2} - \lambda_{FM1} - \lambda_{NM}}{\lambda_{FM1} +\lambda_{FNM}}& \geq 0; \forall p \in [0,1] \\
\frac{\lambda_{FM1}}{\lambda_{FM1} +\lambda_{FNM}} + p\frac{\lambda_{FM2} - \lambda_{FM1} - \lambda_{NM}}{\lambda_{FM1} +\lambda_{FNM}}& \geq \frac{1}{2}; \forall p \in [0,1]
\end{align*}
and thus the sufficient condition is satisfied and $\hat{\mathbf{Z_i}}$ is a bipartite matching and the Bayes estimate of the matching.
\end{proof}
It is also straightforward to show that the conditions in Theorem 1 of \cite{sad2017} imply the conditions of the theorem above. Sadinle's conditions are (1) $\lambda_{FM1} \geq \lambda_{FNM} >0$ and (2) $\lambda_{FM2} \geq \lambda_{FNM} + \lambda_{FM1}$. (1) is the same as the first condition. Under (1), $\frac{\lambda_{FNM} -\lambda_{FM1}}{2}$ can be added to the right hand side of (2) while preserving the inequality, producing the second inequality in Theorem \ref{loss_func}. Thus, the conditions in Theorem \ref{loss_func} are weaker than those of Theorem 1 in \cite{sad2017}. In particular, $\lambda_{FM2}$ is not constrained to be larger than the sum of the other two parameters, or even bigger than $\lambda_{FM1}$.

\section*{Supplementary Materials}
\begin{description}

\item[FBRSRL.R:] R script containing functions, including main function {\tt FBRSRL()}, for constructing comparison data, sampling the posterior and producing a matching. Includes notes on use. (R file)
\item[gibbs\_c.cpp:] C++ code for Step 3 of the Gibbs sampler. To be compiled by {\tt Rcpp} package before running {\tt FBRSRL()}.
\item[gibbs\_c\_asym.cpp:] C$++$ code for Step 3 of the Gibbs sampler when $n_B/n_A >200$. To be compiled by {\tt Rcpp} package before running {\tt FBRSRL()}.

\end{description}

\newpage
\bibliographystyle{apacite}
\bibliography{matchingrefs}

\end{document}